| IPEA panel 2 |
| --- |
| Artificial Intelligence for Climate: Sustainable Infrastructure and Democratic Governance |

## The unsuitability of existing regulations to reach sustainable AI

*Dr. Prof. Thomas Le Goff's contribution to Pre-COP Side Event "From Knowledge to Action: Strategic Dialogue towards COP30", Institute for Applied Economic Research (Ipea) – Brazil, 9-10 October 2025.*

| How do European regulations grasp (or fail to grasp) the issues related to AI sustainability? |
| --- |

Despite some progress in developing rules that address AI systems' sustainability, significant gaps remain. This section will examine the European AI Act (1.), the European regime of environmental, societal and corporate governance (ESG) reporting (2.) and the regulations governing data centers (3.). We argue that, while these measures provide a starting point, they are insufficient to fully address the market failures that constitute the environmental externalities of AI development.

## 1. The limits of sustainability-related provisions in the European AI Act

The European AI Act represents the first binding framework to regulate AI systems in the world and aims to ensure that AI adoption is aligned with European values. While its focus is on safety (the AI Act being a product safety regulation), it also includes some provisions related to environmental sustainability. For instance, the Act imposes documentation obligations on high-risk AI systems[1] and General Purpose AI (GPAI)[2], requiring developers to provide information about their systems' computational resources. Additionally, the AI Act encourages the integration of environmental considerations into voluntary codes of conduct that will be drafted by European standardization organizations and that AI providers can adhere to[3].

However, these provisions are limited in scope and effectiveness. First, the documentation requirements focus narrowly on the computational resources used during the development phase of AI systems, largely neglecting the energy consumed during the inference phase[4]. Since the inference phase often constitutes a significant portion of an AI system's total energy consumption[5], this constitutes a problematic loophole in the AI Act's attempt to promote environmental transparency. Second, the reliance on voluntary codes of conduct means that AI providers can choose whether to adopt environmentally responsible practices, with no binding obligations to do so. This lack of enforceable requirements also creates a loophole that could allow companies to continue prioritizing performance over sustainability, thereby failing to correct the underlying market failures.

---

[1] AI Act, article 11(1) completed by Annex IV.
[2] AI Act, article 53(1) completed by Annex XI.
[3] AI Act, articles 40 and 112.
[4] Ebert, K., Alder, N., Herbrich, R., Hacker, P. 2024. "AI, Climate, and Regulation: From Data Centers to the AI Act". *arXiv*, arXiv:2410.06681, Online: http://arxiv.org/abs/2410.06681.
[5] Luccioni, S., Jernite, Y., Strubell, E. 2024b. "Power Hungry Processing: Watts Driving the Cost of AI Deployment?". *The 2024 ACM Conference on Fairness, Accountability, and Transparency,* 85-99. https://doi.org/10.1145/3630106.3658542.

## 2. The promises of ESG reporting obligations for AI providers and deployers

The legal regime for ESG reporting in the EU represents another potential mechanism for promoting transparency around the environmental impact of AI. The Corporate Sustainability Reporting Directive (CSRD)[6], which expands the scope of the earlier non-financial reporting directive, aims to enhance transparency regarding the sustainability practices of companies operating in the EU. The CSRD requires large companies to disclose detailed information on how their activities affect the environment, including carbon emissions and resource use. As such, AI providers or deployers operating in Europe may fall under the directive's scope and could be expected to report on the environmental impacts of their AI activities. In theory, the CSRD could create incentives for AI companies to adopt more sustainable practices by making their environmental impact more visible to investors, consumers, and other stakeholders[7]. The increased transparency could shift market preferences towards companies that demonstrate a commitment to sustainability, potentially driving competition based on environmental performance. However, the impact of the CSRD will largely depend on how effectively it is enforced. The directive lacks robust enforcement mechanisms, as it imposes no sanctions for non-compliance[8]. Furthermore, the quality and completeness of the information provided through ESG reporting remain uncertain. If the reported data is incomplete or inconsistent, it may not provide stakeholders with the clarity needed to differentiate between companies based on their environmental footprint.

Apart from ESG reporting, a second European directive, the Corporate Sustainability Due Diligence Directive (CSDDD)[9], goes a step further by imposing due diligence obligations on large companies, requiring them to identify, prevent, and mitigate adverse environmental impacts throughout their supply chains[10]. This directive could apply to AI providers or deployers by requiring them to conduct due diligence on the environmental practices of their suppliers, including data center operators or hardware manufacturers. In theory, such obligations could drive companies to adopt more sustainable practices across their value chains, potentially addressing some of the environmental risks associated with AI. However, the CSDDD's ability to foster sustainable AI practices is also limited. While it introduces a due diligence obligation, it does not specifically address the unique environmental challenges posed by AI technologies, such as the energy consumption of large-scale model training. Its focus is on broad supply chain impacts, and as a result, it may not directly target the specific practices that drive the environmental footprint of AI. Additionally, while the directive opens the possibility for civil society and NGOs to hold companies accountable for failing to meet their due diligence obligations[11], further research will be needed to determine whether this mechanism can fully address the liability gaps related to AI's environmental impact, such as failures to meet climate neutrality targets. Thus, while the CSDDD represents a step toward accountability, it does not provide an adapted solution to correct the market failures of AI development's environmental externalities, not mentioning the fact that its application has been postponed by the European Commission as part of the Omnibus 1 Directive[12].

---

[6] Directive (EU) 2022/2464 of the European Parliament and of the Council of 14 December 2022 amending Regulation (EU) No 537/2014, Directive 2004/109/EC, Directive 2006/43/EC and Directive 2013/34/EU, as regards corporate sustainability reporting.
[7] Sætra, H.S. 2021. "A Framework for Evaluating and Disclosing the ESG Related Impacts of AI with the SDGs". *Sustainability*, 13(15): 8503. https://doi.org/10.3390/su13158503.
[8] Bassetti, C. 2024. "Corporate Sustainability Reporting Directive: Timides Progres dans le Domaine du Reporting Extra-Financier". *Sapienza Legal Papers*, 11, 295-312.
[9] Directive (EU) 2024/1760 of the European Parliament and of the Council of 13 June 2024 on corporate sustainability due diligence and amending Directive (EU) 2019/1937 and Regulation (EU) 2023/2859.
[10] Ventura, L. 2023. "Corporate Sustainability Due Diligence and the New Boundaries of the Firms in the European Union". *European Business Law Review*, 34(2): 239-268. https://doi.org/10.54648/eulr2023018.

**3. Addressing the environmental impacts of AI through the regulation of data centers**

Data centers, which serve as the infrastructure backbone for most AI systems, contribute significantly to the environmental footprint of AI through energy consumption, water use, and the demand for materials. As such, they represent an important area for regulation when considering the environmental impact of AI. In the European Union, data centers are subject to various regulatory measures. These include environmental assessment procedures, which require data centers to undergo evaluations of their potential environmental effects before construction, as well as adherence to the EU's Data Center Code of Conduct[13], which sets voluntary guidelines for energy efficiency. Additionally, data centers are regulated through their interconnection processes with electricity grids, which can include requirements for energy sourcing and efficiency improvements[14]. These processes can differ at national level[15].

Despite this framework, the voluntary nature of some of these frameworks means that compliance is not guaranteed, and there is no insurance that data centers will adopt best practices in energy management or cooling techniques. Furthermore, while environmental assessments can help mitigate the impact of new data center construction, they do not address the ongoing environmental impacts of existing facilities, especially those that host large-scale AI models[16]. The focus on site-specific impacts also means that the cumulative effects of the expansion of data centers, driven by the increasing demand for AI compute, are not adequately taken into consideration.

In conclusion, while existing European regulations such as the AI Act, the CSRD, the CSDDD, and data center regulations have the potential to promote transparency and encourage better environmental practices, they fall short of addressing the specific market failures that make AI's environmental footprint a growing concern. The voluntary nature of most provisions, the lack of concrete enforcement mechanisms, and the disconnection between AI-specific impacts and broader sustainability regulations mean that the current regulatory framework is well suited to address the environmental impacts of AI in the long term.

---

[11] Nartey, E.K. 2024. "Addressing corporate human rights violations and environmental harm: advancing holistic remedial framework through tort law and the eu corporate sustainability due diligence directive (CSDDD)". *Athens Journal of Law (AJL)*, 10(3), 345-372.

[12] Proposal postponing the application of some reporting requirements in the CSRD and the transposition deadline and application of the CSDDD - Omnibus I - COM(2025)80 (February 25th).

[13] Acton, M., Bertoldi, P., Booth, J. 2024. "2024 Best Practice Guidelines for the EU Code of Conduct on Data Centre Energy Efficiency". *Joint research center*, JRC136986.

[14] Loeffler, C., Spears, E. 2015. "Uninterruptible Power Supply System". In Geng, H. (ed.) *Data center Handbook*, John Wiley & Sons, pp.495-521.

[15] Le Goff, T., Inderwildi, O., Baldursson, F., Von der Fehr, N.-H. 2025. *From Gridlock to Grid Asset: Data Centres for Digital Sovereignty, Energy Resilience, and Competitiveness*. CERRE report (30th September, 2025). https://cerre.eu/publications/from-gridlock-to-grid-asset-data-centres-for-digital-sovereignty-energy-resilience-and-competitiveness/.

[16] Ebert, K., Alder, N., Herbrich, R., Hacker, P. 2024. "AI, Climate, and Regulation: From Data Centers to the AI Act". *arXiv*, arXiv:2410.06681, Online: http://arxiv.org/abs/2410.06681.

**In the European experience, how the academia/research institutes/think tanks are acting to influence this debate?**

The debate features many different actors, that we could map as follows:

- **Academia**: researchers from various institutions and disciplines work on producing analysis of AI systems' environmental impacts, including Life-Cycle-Assessment[17], critical analysis of regulatory initiatives[18], or developing smaller, frugal and more efficient AI models[19].

- **NGOs and civil society**, including the press, reveal local impacts of AI adoption, specifically regarding data centres. One can find examples in Brazil with revelations about projects putting in danger the rights and the environment of indigenous communities[20] or in France with the work of NGO "La Quadrature du Net" investigating the local impacts of data centers in Marseille[21].

- **International initiatives**, like the Sustainable AI Coalition, promote sustainable practices in AI, officially launched during the Paris AI Action Summit in February 2025.

- **Think tanks** in Europe publish reports on AI sustainability mainly with business-friendly orientations and focus on energy-related topics[22]. Think tanks are subject to private capture by the industry (including big tech companies), so their reports prove to be more in favour of soft regulation to promote innovation, in support to the European Commission's objective to make of the EU a global leader in AI.

- **Standardization bodies** like CEN/CENELEC in Europe or AFNOR in France organize collaborations to produce frugal AI standard, with the will to develop international standards with institutions like ITU, building on existing initiatives. AFNOR has published the first general framework on frugal AI, identifying 31 best practices to minimize the environmental impacts of an AI project[23]. Standardization bodies are also subject to a risk of private capture by the industry.

---

[17] Sophia Falk, David Ekchajzer, Thibault Pirson, Etienne Lees-Perasso, Augustin Wattiez, Lisa Biber-Freudenberger, Sasha Luccioni, Aimee van Wynsberghe, More than Carbon: Cradle-to-Grave environmental impacts of GenAI training on the Nvidia A100 GPU". 2025. *arXiv*, 2509.00093. https://doi.org/10.48550/arXiv.2509.00093.

[18] Hacker, Philipp. 2024. "Sustainable AI Regulation". *Common Market Law Review* 61(2): 345-386. https://doi.org/10.54648/cola2024025.

[19] See the work of Dr. Loïc Lannelongue (https://www.lannelongue-group.org) or Dr. Prof. Anne-Laure Ligozat (https://cv.hal.science/anne-laure-ligozat).

[20] Martins, L. 2025. "Indigenous group in Brazil takes TikTok to court over planned data center". *Rest of world* (8th September 2025). https://restofworld.org/2025/brazil-indigenous-group-sues-tiktok-data-center/.

[21] La Quadrature du Net. 2024. "Enquête : à marseille comme ailleurs, l'accaparement du territoire par les infrastructures du numérique". Online (20th November 2024): https://www.laquadrature.net/2024/11/20/accaparement-du-territoire-par-les-infrastructures-du-numerique/.

[22] See Brueghel's work (https://www.bruegel.org/event/ai-and-energy-sector-navigating-two-way-transformation) or CERRE report on Data Centers and electricity systems in Europe : CERRE. 2025. From Gridlock to Grid Asset: Data Centres for Digital Sovereignty, Energy Resilience, and Competitiveness. Report (30th September 2025). https://cerre.eu/publications/from-gridlock-to-grid-asset-data-centres-for-digital-sovereignty-energy-resilience-and-competitiveness/.

[23] https://telechargement.afnor.info/standardization-afnor-spec-ai-frugal

- **Industry**, mainly tech companies, focuses on developing new techniques and methodologies to advance the efficiency of AI systems. Examples include Nvidia putting efforts on developing more efficient chips and GPUs[24] or Google selling products to help the development of solar panels[25]. Very few companies are transparent on the comprehensive environmental impacts of their products, most of them are opposed to any regulatory interventions. First initiatives like the life-cycle-analyses published by Mistal AI[26] and Google on Gemini[27] need to be highlighted and prove that public pressure succeeds in making business practices more transparent.

**What strategic recommendations would you propose for implementing the COP30 Action Agenda?**

1. **Incorporate environmental sustainability requirements – including transparency over carbon emissions, water use and material needs – in global AI regulatory frameworks**.

2. As one can reduce what one can measure, **promote environmental transparency from AI providers**.

3. **Commit to develop international standards on AI sustainability**, including methodologies for AI systems' lifecycle assessments, common metrics for environmental transparency (energy needs, carbon emissions, water use, material needs), and methodologies to choose the right AI model for the right task, promoting the use of more frugal AI models and striking a balance between technical and environmental performance.

4. **Ensure the promotion of data centres does not lead to uncontrolled adoption and disproportionate environmental impacts or fundamental rights infringement**. **Ensure all projects conduct an environmental impact assessment** and adopt a mitigation plan to address the risks identified.

5. Following the international principle public participation in environmental decision-making (UN Rio Declaration, principle 10), **ensure the involvement of the public and local communities in decision-making about AI infrastructures, by giving them transparent information and a voice in the debate**.

---

[24] https://images.nvidia.com/aem-dam/Solutions/documents/FY2024-NVIDIA-Corporate-Sustainability-Report.pdf
[25] https://ai.google/sustainability/
[26] https://mistral.ai/news/our-contribution-to-a-global-environmental-standard-for-ai
[27] https://cloud.google.com/blog/products/infrastructure/measuring-the-environmental-impact-of-ai-inference?hl=en